\documentclass[sigconf]{acmart}
\AtBeginDocument{%
  }

\copyrightyear{2025}
\acmYear{2025}
\setcopyright{cc}
\setcctype{by}
\acmConference[MM '25]{Proceedings of the 33rd ACM International Conference on Multimedia}{October 27--31, 2025}{Dublin, Ireland}
\acmBooktitle{Proceedings of the 33rd ACM International Conference on Multimedia (MM '25), October 27--31, 2025, Dublin, Ireland}
\acmDOI{10.1145/3746027.3755233}
\acmISBN{979-8-4007-2035-2/2025/10}

\definecolor{best1}{RGB}{222,242,212}
\definecolor{best2}{RGB}{255,250,212}

\begin{document}

\title{BEAM: Bridging Physically-based Rendering and Gaussian Modeling for
Relightable Volumetric Video}


\author{Yu Hong}
\authornote{Both authors contributed equally to this research.}
\orcid{0009-0004-1831-7652}
\affiliation{%
  \institution{ShanghaiTech University}
  \city{Shanghai}
  \country{China}
}
\affiliation{%
  \institution{Neudim}
  \city{Shanghai}
  \country{China}
}
\email{hongyu@shanghaitech.edu.cn}

\author{Yize Wu}
\authornotemark[1]
\orcid{0009-0004-0131-114X}
\affiliation{%
  \institution{ShanghaiTech University}
  \city{Shanghai}
  \country{China}
}
\email{wuyize@shanghaitech.edu.cn}

\author{Zhehao Shen}
\affiliation{%
  \institution{ShanghaiTech University}
  \city{Shanghai}
  \country{China}}
\affiliation{%
  \institution{Neudim}
  \city{Shanghai}
  \country{China}
}
\orcid{0009-0001-8933-0385}
\email{shenzhh@shanghaitech.edu.cn}

\author{Chengcheng Guo}
\affiliation{%
  \institution{ShanghaiTech University}
  \city{Shanghai}
  \country{China}}
\orcid{0009-0003-2122-7001}
\email{guochch2024@shanghaitech.edu.cn}

\author{Yuheng Jiang}
\orcid{0000-0001-8121-0015}
\affiliation{%
 \institution{ByteDance Inc.}
 \city{Shanghai}
 \country{China}}
\email{nowheretrix123@gmail.com}

\author{Yingliang Zhang}
\affiliation{%
  \institution{DGene}
  \city{Shanghai}
  \country{China}}
\email{yingliang.zhang@dgene.com}
\orcid{0000-0002-0594-7549}

\author{Qiang Hu}
\affiliation{%
  \institution{Shanghai Jiao Tong University}
  \city{Shanghai}
  \country{China}}
\email{qiang.hu@sjtu.edu.cn}
\orcid{0000-0003-4645-9776}
\authornote{Corresponding Author.}

\author{Jingyi Yu}
\orcid{0000-0002-8580-0036}
\affiliation{%
  \institution{ShanghaiTech University}
  \city{Shanghai}
  \country{China}}
\email{yujingyi@shanghaitech.edu.cn}

\author{Lan Xu}
\affiliation{%
  \institution{ShanghaiTech University}
  \city{Shanghai}
  \country{China}}
\orcid{0000-0002-8807-7787}
\email{xulan1@shanghaitech.edu.cn}
\authornotemark[2]

\renewcommand{\shortauthors}{Yu Hong et al.}

\begin{abstract}
Volumetric video enables immersive experiences by capturing dynamic 3D scenes, enabling diverse applications for virtual reality, education, and telepresence. However, traditional methods struggle with fixed lighting conditions, while neural approaches face trade-offs in efficiency, quality, or adaptability for relightable scenarios. To address these limitations, we present BEAM, a novel pipeline that bridges 4D Gaussian representations with physically-based rendering (PBR) to produce high-quality, relightable volumetric videos from multi-view RGB footage. BEAM recovers detailed geometry and PBR properties via a series of available Gaussian-based techniques. It first combines Gaussian-based human performance tracking with geometry-aware rasterization in a coarse-to-fine optimization framework to recover spatially and temporally consistent geometries. We further enhance Gaussian attributes by incorporating PBR properties step by step. We generate roughness via a multi-view-conditioned diffusion model, and then derive AO and base color using a 2D-to-3D strategy, incorporating a tailored Gaussian-based ray tracer for efficient visibility computation. Once recovered, these dynamic, relightable assets integrate seamlessly into traditional CG pipelines, supporting real-time rendering with deferred shading and offline rendering with ray tracing. By offering realistic, lifelike visualizations under diverse lighting conditions, BEAM opens new possibilities for interactive entertainment, storytelling, and creative visualization.
\end{abstract}




\maketitle

\begin{figure}[h]
  \includegraphics[width=\columnwidth]{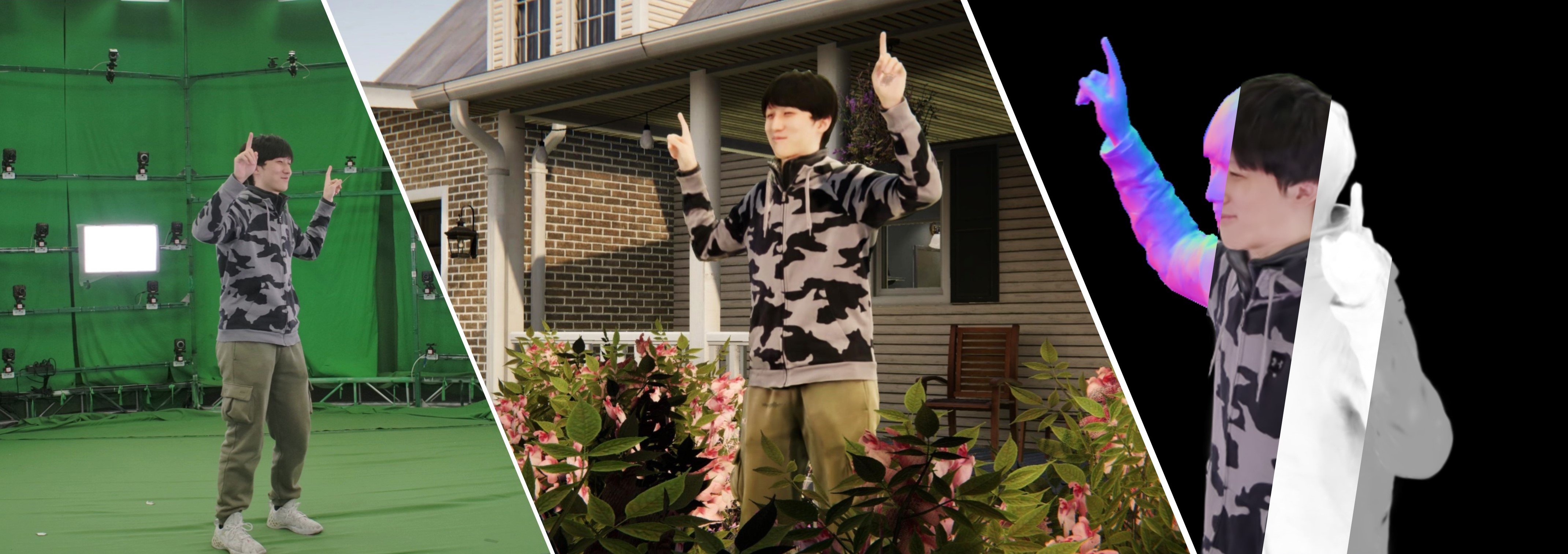}
  \captionof{figure}{We present BEAM, a novel pipeline that bridges 4D Gaussians with accurate physically-based rendering to produce relightable volumetric videos, delivering immersive and realistic experiences on platforms such as VR.}
  \label{fig:teaser}
\end{figure}

\section{Introduction}
\label{sec:1_intro}
Volumetric video captures dynamic 3D scenes from multiple angles, allowing interactive viewing from any perspective. 
This technology is crucial for creating immersive experiences in virtual and augmented reality, enhancing storytelling, education, cultural preservation, and telepresence with lifelike, interactive content. However, traditional volumetric video is often limited by fixed lighting conditions captured during recording, which can clash with dynamic or virtual environments, reducing realism and flexibility. Relightable volumetric video overcomes this limitation by enabling post-capture relighting. This allows for seamless integration into dynamic lighting environments and offers creative control over visual aesthetics.

The prevailing workflow~\cite{collet2015high, dou2016fusion4d, lawrence2024project, garon2016real} for producing relightable volumetric videos in the industry still relies on tracked mesh sequences and texture videos, which can be seamlessly integrated into standard CG pipelines to support relighting under various lighting conditions. However, the intricate reconstruction process often introduces artifacts such as holes and noise, and the quality of relighting remains constrained, frequently resulting in visible imperfections.
Neural advancements~\cite{srinivasan2021nerv, zhang2021nerfactor, yao2022neilf} focus on enabling relighting capabilities using neural factorization within implicit MLPs representations. However, these approaches often face challenges in balancing training efficiency, rendering speed, and output quality, ultimately failing to deliver satisfactory results.
Recently, 3D Gaussian Splatting (3DGS)~\cite{gaussiansplatting}, an efficient point-based representation, has achieved photo-realistic rendering at unprecedented frame rates. While dynamic variants~\cite{jiang2024robust, li2023spacetime, vcube} can produce high-quality volumetric videos, they fail to produce the detailed geometry necessary for essential operations like relighting.
Although efforts~\cite{gao2025relightable, shi2023gir, liang2024gs, jiang2024gaussianshader} have been made to integrate physically-based rendering into the 3DGS pipeline, these methods are often computationally expensive and limited to static scenarios. These limitations severely restrict their applicability in industrial workflows, hindering the efficient production of 4D content.

In this paper, we introduce BEAM, a novel pipeline that bridges 4D Gaussians with accurate physically-based rendering (PBR) for producing relightable volumetric videos from multi-view RGB footage. Our key idea is to robustly recover detailed geometry of human performances and decouple the PBR properties (e.g., ambient occlusion, roughness, and base color) using a carefully selected suite of techniques, i.e., rasterization, performance tracking, and ray tracing, all within a Gaussian-based paradigm. As a result, BEAM enables lifelike dynamic scenes that can be seamlessly and CG-friendly integrated into various platforms under diverse lighting (see Fig.~\ref{fig:teaser}). 

We first recover detailed and spatial-temporally consistent geometries from multi-view video input, which organically combines the Gaussian-based performance tracking~\cite{jiang2024robust} with the geometry-aware Gaussian rasterization~\cite{zhang2024rade}. While the former excels at motion tracking and the latter at static geometry recovery, we unify them in a coarse-to-fine optimization framework. Specifically, we employ coarse joint Gaussians to track non-rigid motion and dense skin Gaussians to preserve intricate geometry details. We adopt a robust optimization process that integrates normal consistency, photometric consistency, and temporal regularization to enhance geometric accuracy and smoothness. This enables accurate depth and normal recovery from the dense Gaussians using the geometry-aware rasterizer~\cite{zhang2024rade}, providing a robust foundation for material decomposition and relighting.

We further decouple the dense 4D Gaussians to recover detailed material properties, enabling high-quality physically-based rendering grounded in the rendering equation~\cite{kajiya1986rendering} and simplified Disney BRDF~\cite{burley2012physically}.
Assuming human-centric scenes with negligible metallic components, we focus on accurately associating roughness, ambient occlusion (AO), and base color properties with the Gaussians, ensuring realistic and adaptable rendering under diverse lighting conditions.
To achieve this, we adopt a step-by-step approach to disentangle these properties.  
Specifically, we first generate a roughness texture using the material diffusion module in previous work~\cite{zhang2024clay} with multi-view conditioning, which is associated with the dense Gaussians through UV projection. Then, for the AO and base color, we adopt a 2D-to-3D strategy, where these attributes are estimated in the input views to bake 2D material maps, and then optimized into the corresponding dense Gaussians in the 3D space. This strategy effectively reduces noise and smooths the disentanglement to improve relighting quality. For further 2D AO and base color decomposition, the lighting environment during capturing can be estimated using an off-the-shelf tool~\cite{new_house_internet_services_bv_ptgui_2025}, while the geometry attributes and roughness are obtained in previous stages. Thus, by carefully re-examining and simplifying the rendering equation~\cite{kajiya1986rendering}, we identify a critical insight: both 2D AO and base color can be accurately derived by accumulated visibility information for specific points along specific directions during ray tracing. We tailor the Gaussian-based ray tracer~\cite{3dgrt2024} to compute such visibility, with a novel alpha blending strategy based on the dense Gaussians. This strategy efficiently captures visibility information, forming the foundation for estimating AO and base color maps in the input viewpoints.

\begin{figure*} [ht]
  \label{fig:pipeline}
  \centering
  \includegraphics[width=\textwidth]{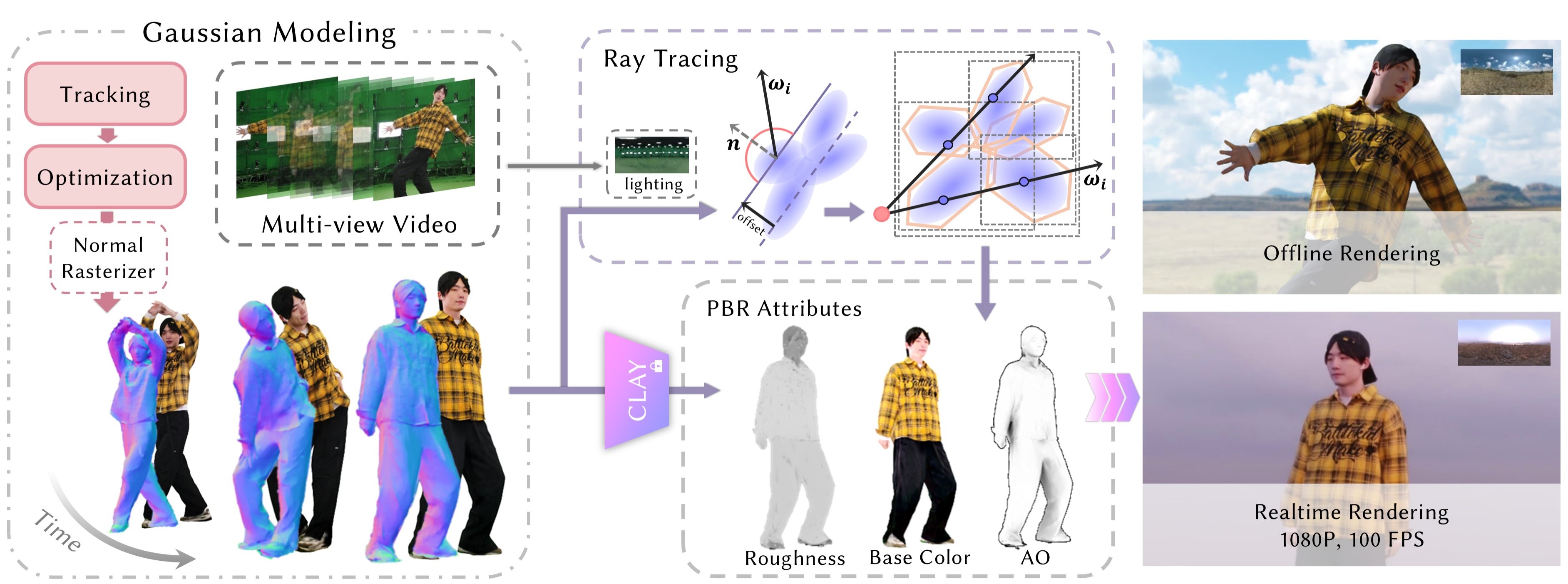}
  \captionof{figure}{We propose a novel BEAM method to produce Gaussian sequences. We first use tracking results of joint Gaussians and a normal regularizer to obtain our 4D Gaussians with consistent geometry. Then we infer roughness using a generative model and apply ray tracing to compute the 2D base color and AO maps, which are then used to optimize the corresponding Gaussian attributes. Our results can be rendered under varying lighting conditions using both real-time and offline rendering.} 
  \label{fig_pipeline}
  \vspace{-4pt}
\end{figure*}

Once the material properties are baked into our dense dynamic Gaussians, these 4D assets seamlessly integrate with traditional CG engines, supporting flexible rendering workflows. For real-time rendering, we adopt deferred shading to deliver immersive and efficient visualizations, while offline rendering leverages ray tracing to precisely capture complex shadows and occlusions.
We further develop a Unity plugin enabling seamless integration of 4D assets into various platforms for real-time, lifelike interactions under diverse lighting conditions. This innovation opens new possibilities for storytelling, interactive entertainment, and creative visualization, offering users an immersive journey into dynamic, relightable volumetric worlds.

\vspace{-20pt}

\section{Related Work}
\label{sec:related}

\paragraph{Human Modeling}

In the field of Human Modeling, numerous methods~\cite{sun2021neural,suo2021neuralhumanfvv,fridovich2023k,hu2025varfvv} have been proposed to address these challenges. Li~\cite{Temporally} integrates temporal denoising into non-rigid mesh template tracking to capture detailed geometry, while High-quality FVV~\cite{collet2015high} compactly represents human performance using tracked mesh sequences and textured video.

Building on DynamicFusion~\cite{newcombe2015dynamicfusion}, subsequent methods have integrated techniques like SIFT features~\cite{innmann2016volumedeform}, motion field constraints~\cite{slavcheva2017killingfusion,slavcheva2018sobolevfusion}, neural deformation graphs~\cite{bozic2021neural}, and parametric human models like SMPL~\cite{loper2023smpl} to address challenges in human reconstruction, including drift and human-object interactions. 

With the advancement of neural rendering techniques~\cite{nerf}, several works have incorporated time as a latent variable to handle dynamic scenes~\cite{pumarola2021d,xian2021space,tretschk2021non}. Meanwhile, other approaches~\cite{peng2021neural,zheng2023avatarrex,habermann2023hdhumans,sun2024real,zhu2024trihuman} leverage the human parametric model~\cite{loper2023smpl} to reconstruct the animatable avatar. Furthermore, some hybrid methods~\cite{yu2021function4d, jiang2022neuralhofusion,jiang2023instant} combine explicit volumetric fusion with implicit neural techniques to capture more details. 
Additionally, several methods~\cite{wang2023neural,icsik2023humanrf,shao2023tensor4d,hu2025vrvvc} significantly accelerate training and rendering speed, by leveraging advanced data structures such as voxel grids~\cite{fridovich2022plenoxels}, hash tables~\cite{muller2022instant}, and tensor decomposition~\cite{chen2022tensorf}.
Recently, 3DGS~\cite{gaussiansplatting} ensures both high quality and fast rendering, while dynamic variants~\cite{sun20243dgstream,luiten2024dynamic,huang2024sc,yang20244d,jiang2024hifi4g,duan20244d,hu20254dgc,li2023spacetime,jiang2025reperformer} enable complex 4D scene reconstruction for advanced human modeling. DualGS~\cite{jiang2024robust} uses joint and skin Gaussians to capture motion and detailed appearance. However, these approaches fail to recover detailed geometry and do not support relighting.

\paragraph{Human Relighting}
Human Relighting aims to manipulate the reflectance field of the human surface, enabling an immersive fusion with novel illumination. Conventional methods~\cite{chabert2006relighting,debevec2012light,guo2019relightables,hawkins2001photometric,wenger2005performance,weyrich2006analysis} use LightStage systems to capture human reflectance characteristics, requiring costly controlled lighting and dense camera arrays that are not widely accessible. In 2D image-based relighting tasks, previous methods~\cite{Kanamori_Endo_2018, Tajima_Kanamori_Endo_2021} rely on convolutional networks for inference, while recent diffusion-based approaches~\cite{ding2023diffusionriglearningpersonalizedpriors, Zeng_2024} leverage generative models to learn complex light interactions across large datasets. However, the lack of a 3D representation makes it challenging to maintain 3D consistency of lighting. In addition, some neural relighting methods~\cite{boss2021nerd,zhang2021nerfactor,zeng2023relighting} use NeRF~\cite{nerf} for 3D reconstruction of material properties and lighting effects. However, their rendering quality is inadequate and hard to integrate with traditional CG pipelines.
Recent methods~\cite{gao2025relightable,jiang2024gaussianshader,liang2024gs, gu2024irgs} leverages 3DGS representation for relighting due to its ability to reconstruct fine details and interaction in CG engines. 
For human performance relighting, researchers~\cite{chen2022relighting4d, xu2024relightable, li2024animatablegaussians, chen2024meshavatar, zheng2025physavatar, luvizon2023relightable} extend mesh-based and neural relighting methods by incorporating body pose priors~\cite{loper2023smpl, jiang2024smplx}. However, avatars relying on skeletal priors struggle with complex clothing, wrinkles, and human-object interactions.
\section{Method}
\label{sec:method}

Given multi-view video inputs with known environment illumination, we aim to produce relightable 4D Gaussian sequences with physically-based rendering (PBR) materials, enabling realistic rendering under diverse lighting conditions. PBR typically requires multiple material components, such as base color, metallic, roughness, normal, and ambient occlusion (AO). As the metallic attribute is negligible for human bodies, we set it to zero and focus on optimizing the remaining properties: normal, roughness, AO, and base color. 
The resulting relightable 4D Gaussians are rendered under various lighting and scene configurations, and further integrated into CG engines and VR applications. The complete pipeline is illustrated in Fig.~\ref{fig_pipeline}.

\subsection{Gaussian Modeling and Geometry Optimization}
\label{sec_3_1}

To obtain temporally consistent geometric information (depth and normal maps) for physically-based rendering, we seamlessly combine the Gaussian-based performance tracking~\cite{jiang2024robust} with the geometry-aware rasterizer~\cite{zhang2024rade} within a coarse-to-fine optimization framework. Specifically, our framework uses a dual Gaussian representation to separately model global motion and visual appearance through joint and skin Gaussians. Each skin Gaussian is anchored to multiple joint Gaussians and is warped across frames based on the tracking results of these joint Gaussians. The optimization process integrates a photometric loss $E_{\text{color}}$, a smooth term $E_{\text{smooth}}$, and a temporal regularization term $E_{\text{temp}}$. 

To further enhance the optimization of Gaussian geometry, we introduce an additional normal consistency loss $E_{\text{normal}}$. During rasterization, we assume that the intersection points between rays and Gaussians correspond to the maxima in Gaussian values. The depth of a Gaussian is defined as the depth of the intersection point, while the normal of the intersection plane is taken as the Gaussian normal. This rasterizing process generates precise depth maps and normal maps $N_r$. Using the obtained depth map, we compute the normal map $N_d$ based on a local plane assumption and then measure the normal consistency loss $E_{\text{normal}}$ between $N_d$ and $N_r$ as follows:
\begin{equation}
E_{\text{normal}} = \sum_i \omega_i (1 - N_r^\top N_d),
\end{equation}
where $i$ indexes the intersected splats along the ray, and $\omega_i=\alpha_i \prod_{j=1}^{i-1} (1 - \alpha_j)$ represents the blending weight of the intersection point.

Together, the total energy term in our 4D Gaussian modeling and geometric optimization framework is expressed as:
\begin{equation}
    E = \lambda_{\text{color}}E_{\text{color}}+\lambda_{\text{smooth}}E_{\text{smooth}}+\lambda_{\text{temp}}E_{\text{temp}}+\lambda_{\text{normal}}E_{\text{normal}},
\end{equation}
During dynamic training, $E_{\text{norm}}$ is introduced after the appearance optimization over 7,000 iterations, followed by an additional 5,000 iterations dedicated to optimizing the normals. For other energy terms not discussed here, please refer to DualGS~\cite{jiang2024robust}. The hyperparameters are set as follows: $\lambda_{\text{color}} = 1$, $ \lambda_{\text{smooth}} = 0.001$, $\lambda_{\text{temp}} = 0.00005$, $\lambda_{\text{normal}} = 0.03$.

With the above energy term, we obtain a Gaussian sequence with temporally consistent geometry. For further mesh extraction, we render depth maps for training views and then adopt TSDF~\cite{curless1996volumetric} fusion. The resulting mesh sequence is then used for subsequent PBR material decomposition.

\begin{figure*} [ht]
  \centering
  \includegraphics[width=\textwidth]
  {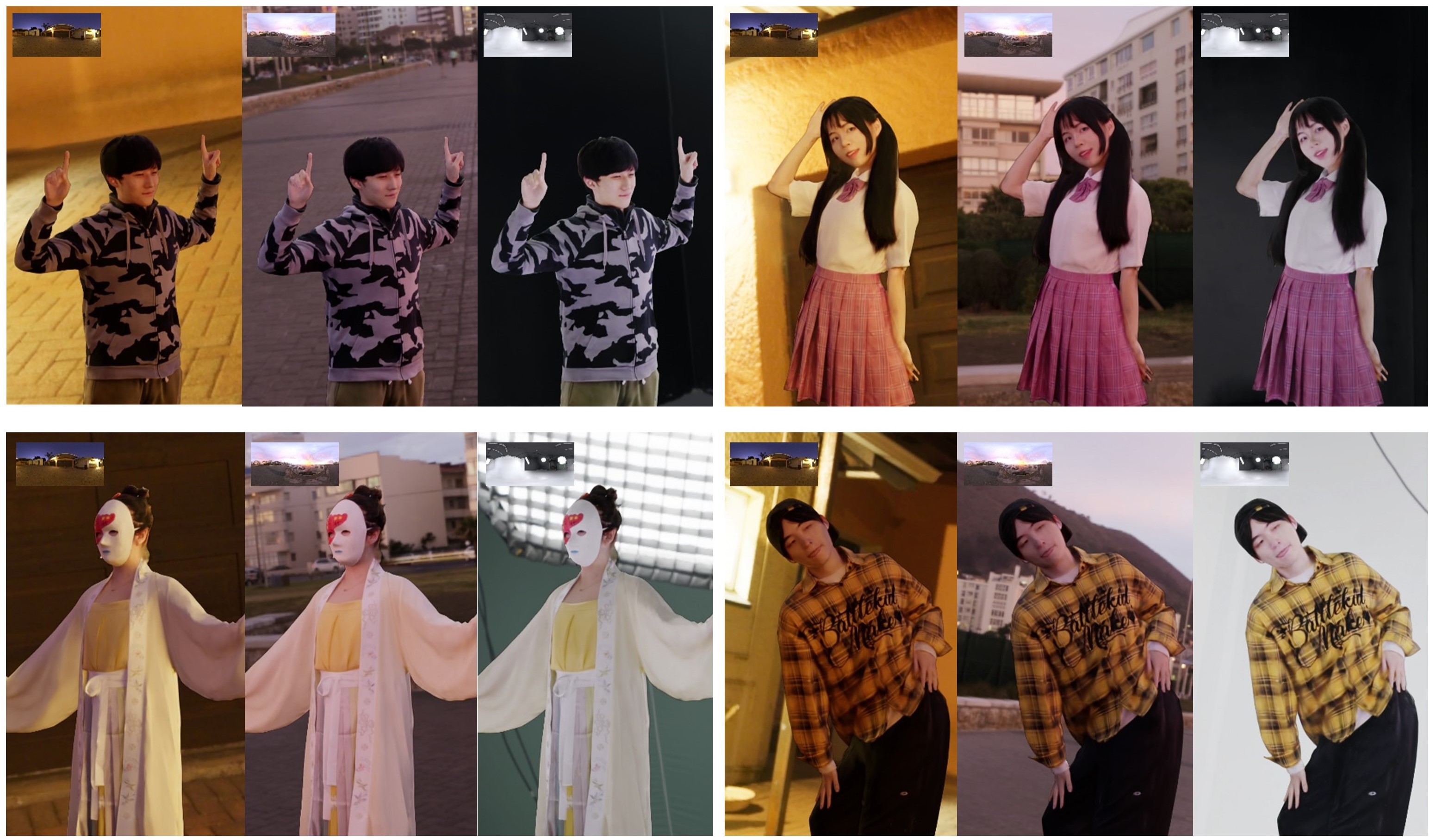}
  \captionof{figure}{Gallery of our results. We present some real-time rendering results under HDRI settings, which deliver high-fidelity rendering of human performances across challenging motions and complex clothing textures.} 
  \vspace{-12pt}
  \label{fig_moreres}
\end{figure*}

\subsection{PBR Materials Decomposition}
\label{sec_3_2}
To enable physically-based rendering within the Gaussian-based paradigm, we enhance the 3D Gaussian representation by introducing additional material attributes: roughness $r \in [0,1]$, ambient occlusion (AO) $\mathcal{A} \in [0,1]$, and base color $\boldsymbol{\rho} \in [0,1]^3$. The $i^{th}$ extended Gaussian $\mathcal{P}_i(t)$ at time $t$ is defined as:
\begin{equation}
    \mathcal{P}_i(t) = \{\boldsymbol{\mu}_i(t), \boldsymbol{q}_i(t), \boldsymbol{s}_i(t), o_i(t), \boldsymbol{c}_i(t), \boldsymbol{\rho}_i(t), r_i(t), \mathcal{A}_i(t)\},
\end{equation}
where each component denotes position, orientation, scale, opacity, color, base color, roughness, and ambient occlusion, respectively.
To better capture subtle appearance changes over time, we allow the base color $\rho$ to vary for each Gaussian. 
The roughness $r$, on the other hand, is kept time-invariant as a stable material property. To disentangle these material properties, we adopt a step-by-step optimization strategy.

\paragraph{Roughness.} 
To efficiently assign a roughness value to each Gaussian, we utilize a generative model that offers significantly faster performance compared to inverse rendering methods. Specifically, we feed the canonical mesh obtained from geometric optimization and multi-view images into the Material Diffusion module of CLAY~\cite{zhang2024clay} to generate a roughness texture for the mesh. Each valid pixel of the texture is then mapped back to the world coordinate system, allowing us to assign each Gaussian the roughness value of its nearest pixel.

\paragraph{Ambient Occlusion and Base Color}  
For AO and base color, we first estimate these two attributes in the input views for each frame to bake 2D material maps, and then optimize into the corresponding dense skin Gaussians.

Ambient Occlusion $\mathcal{A}(x)$ is an approximation of global illumination, which models the diffuse shadows produced by close, potentially small occluders within a constrained computational budget:
\begin{equation}
\label{equ_AO}
\mathcal{A}(x) = \frac{1}{\pi} \int_{\Omega} V^{\text{env}}\left(x,\boldsymbol{\omega}_{i}\right) (\mathbf{n} \cdot \boldsymbol{\omega}_{i}) \mathrm{d} \boldsymbol{\omega}_{i},
\end{equation}
where $V^{\text{env}}\left(x,\boldsymbol{\omega}_{i}\right)$ is the visibility term at 3D point $x$ in direction $\boldsymbol{\omega}_i$, $\mathbf{n}$ is the normal of the surface at point $x$, $\Omega$ is the hemisphere centered in $x$ and having $\mathbf{n}$ as its axis.

For the base color, we use a simplified Disney BRDF model~\cite{burley2012physically} composed of a Lambertian diffuse term and a Cook-Torrance specular term~\cite{cook1982reflectance}, with the outgoing radiance $L_{o}$ being a linear combination of these two components. Since for dielectric materials the diffuse part is proportional to the base color while the specular part is independent of it, the rendering equation~\cite{kajiya1986rendering} can be written as: 
\begin{equation}
\label{equ_Lo}
    L_o\left(x,\boldsymbol{\omega}_{o}\right) = \rho(x) L_{o}^D\left(x,\boldsymbol{\omega}_{o}\right) + L_{o}^S\left(x,\boldsymbol{\omega}_{o}\right),
\end{equation}
where $L_o\left(x,\boldsymbol{\omega}_{o}\right)$ is the outgoing radiance at $x$ in direction $\boldsymbol{\omega}_{o}$, computed by mapping image pixel colors to linear space. $\rho$ is the base color, $L_{o}^D$ is the diffuse part residue and $L_{o}^S$ is the specular part.
To simplify the rendering equation, our computation of the base color disregards the indirect illumination effects caused by surface reflections on the human body, and assumes the environment illumination is distant. Consequently, $L_{o}^D$ and $L_{o}^S$ are expressed as follows:
\begin{align}
\label{equ_LoD}
     L_{o}^D\left(x,\boldsymbol{\omega}_{o}\right) &= \frac{1}{\pi  } \int_\Omega 
     \left(1-F\right)  \mathcal{L}(x,\boldsymbol{\omega}_i) \mathrm{d} \boldsymbol{\omega}_{i},  \\
    \label{equ_LoS}
     L_{o}^S\left(x,\boldsymbol{\omega}_{o}\right) &= \int_\Omega f_{r_s}\left(x,\boldsymbol{\omega}_{i}, \boldsymbol{\omega}_{o}\right)  \mathcal{L}(x,\boldsymbol{\omega}_i) \mathrm{d} \boldsymbol{\omega}_{i},
\end{align}
where $\mathcal{L}(x,\boldsymbol{\omega}_i) = V^{\text{env}}\left(x,\boldsymbol{\omega}_{i}\right) L_{i}^{\text{env}}\left(\boldsymbol{\omega}_{i}\right) (\mathbf{n} \cdot \boldsymbol{\omega}_{i})$, $F$ is the approximated Fresnel term, $f_{r_s}$ is the specular term in the BRDF, $L_{i}^{\text{env}}$ can be queried from the environment map. To capture the environment map of our multi-view, well-lit dome setup, we position a DSLR camera at the center and take bracketed exposure photographs from multiple directions. These photographs are then processed using PTGui~\cite{new_house_internet_services_bv_ptgui_2025} to generate a high dynamic range (HDR) panoramic image, enabling precise calculation of the incoming radiance $L_i^{\text{env}}$ from the environment.

We observe that both 2D AO and base color need the visibility term $V^{\text{env}}$. To accurately compute $V^{\text{env}}$, we 
adopt Gaussian ray tracer from 3DGRT~\cite{3dgrt2024} and compute intersections based on the maximum Gaussian response.  For efficient computation, a proxy icosahedron mesh is employed to leverage hardware acceleration, with a two-level BVH constructed at both the mesh and instance levels. The $V^{\text{env}}$ can be computed from the Gaussians' transmittance along the ray without sorting them:
\begin{equation}
    V^{\text{env}}\left(x,\boldsymbol{\omega}_{i}\right) = \prod_{i=1}^{N} (1 - o_i G_i(x_i)), 
\end{equation}
where $o_i$ is the opacity of the $i^{th}$ Gaussian, $G_i(x_i)$ is the response of the Gaussian kernel at the intersection point $x_i$. Ray tracing is early stopped if the attenuating $V^{\text{env}}$ is less than a threshold $T$ (we choose $T$ as 0.0001).
To avoid occlusion from Gaussians directly above the original depth, we offset the ray origin $o$ along the normal direction: $o=x+\epsilon \mathbf{n}$.

\begin{figure*}[ht]
  \centering
  \includegraphics[width=\textwidth]{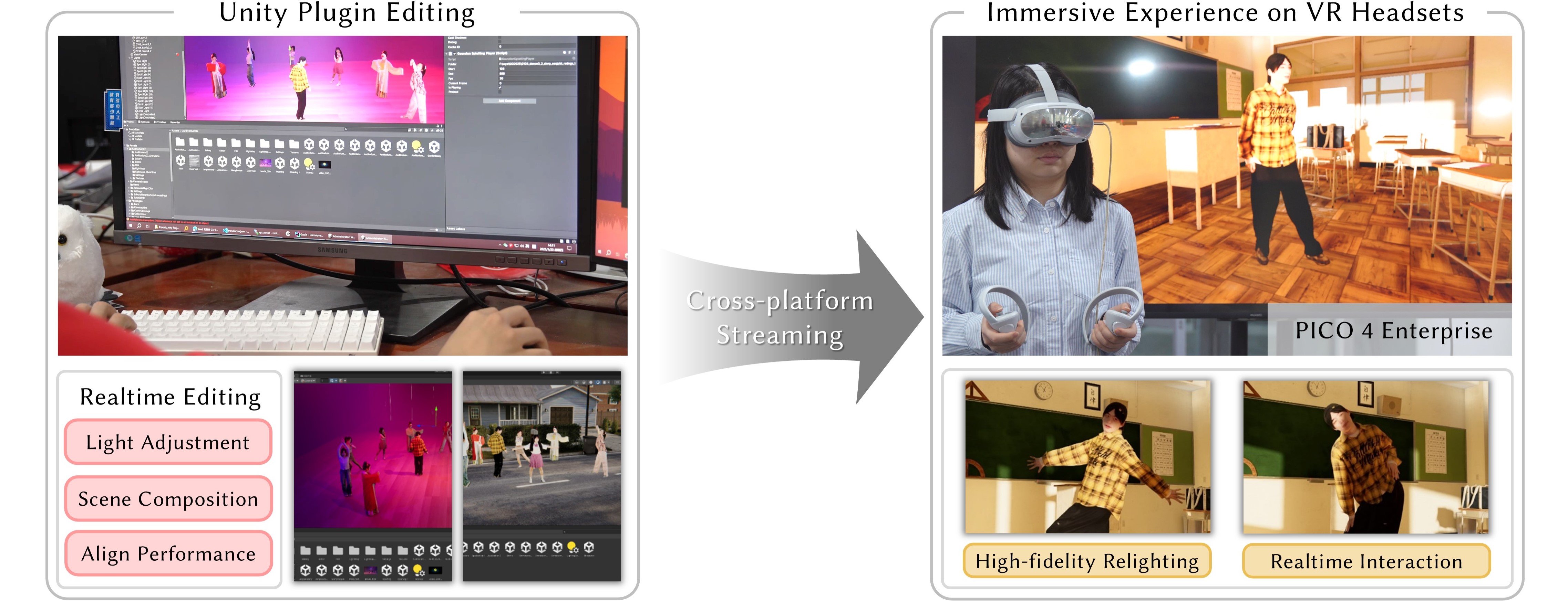}
  \captionof{figure}{We demonstrate scene and lighting editing within Unity, and immersive viewing using VR headsets.} 
  \vspace{-6pt}
  \label{fig_application}
\end{figure*}

\begin{figure}[t]
  \includegraphics[width=\columnwidth]{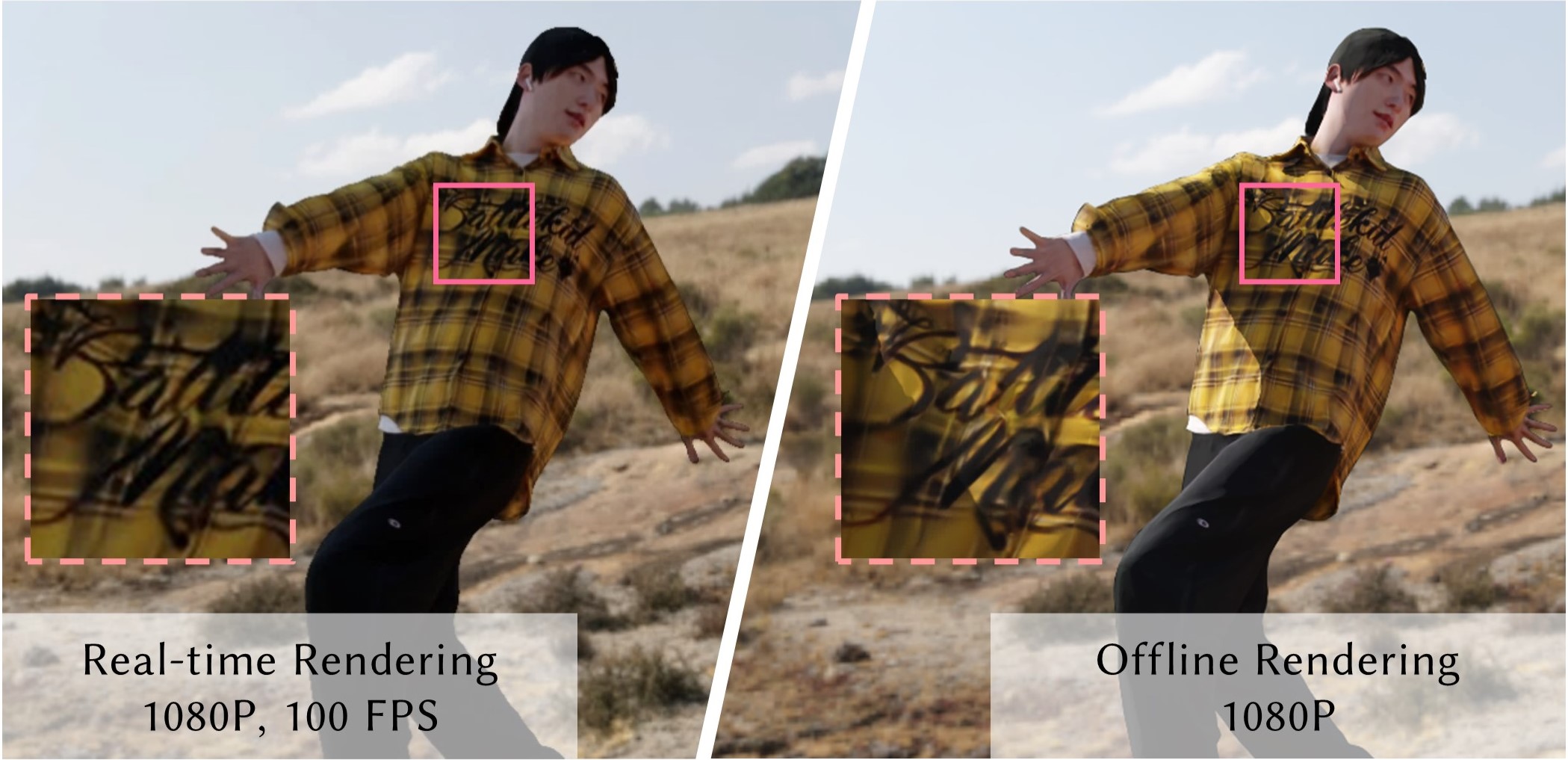}
  \captionof{figure}{We present the results of different rendering techniques. Real-time rendering focuses on efficiency,  while offline rendering delivers more realistic shadows and occlusion effects.} 
  \vspace{-12pt}
  \label{fig_rendering}
\end{figure}

Finally, we use the Monte Carlo method to solve the $\mathcal{A}$ in Eq.~\ref{equ_AO}, $L_o^D$ in Eq.~\ref{equ_LoD}, and $L_o^S$ in Eq.~\ref{equ_LoS}, then the base color $\rho$ can be solved from Eq.~\ref{equ_Lo}. For our Gaussian sampler, we render the AO image with 50 samples per pixel (spp) and the base color image with 100 samples per pixel, then apply Intel Open Image Denoise~\cite{IntelOIDN2025} for denoising. 

To optimize the AO and base color material maps into the corresponding dense skin Gaussians, we employ the generated training view material maps as ground truth supervision while keeping all other geometry-related attributes fixed.
The optimization process commences by initializing the AO attributes with zeros and the base color attributes with RGB values. We then train 5000 iterations to obtain both AO and base color attributes separately. Besides, an $L_2$ regularization term is applied to the base color to ensure temporal consistency.

During optimization, the view-dependent RGB color $\boldsymbol{c}$ is used for material decomposition, but it is no longer required for relighting. After optimization, we discard this attribute and reparameterize each Gaussian as
\begin{equation}
    \mathcal{P}'_i(t) = \{\boldsymbol{\mu}_i(t), \boldsymbol{q}_i(t), \boldsymbol{s}_i(t), o_i(t), \boldsymbol{\rho}_i(t), r_i(t), \mathcal{A}_i(t)\}.
  \vspace{-4pt}
\end{equation}
This compact representation reduces storage and rendering load in CG engines and VR platforms, as $\boldsymbol{c}$ requires high-order spherical harmonics (SH) encoding to capture anisotropic effects.

\subsection{Physically Based Rendering}
\label{sec_3_3}

By leveraging our relightable 4D Gaussians, we can seamlessly integrate the 4D assets into traditional CG engines, supporting both real-time and offline rendering workflows.
For real-time rendering, we utilize deferred shading to deliver immersive and efficient visualization across diverse settings. For offline rendering, we employ ray tracing, which excels in handling shadows and occlusion relationships, ensuring high-quality results.
\paragraph{Real-time Rendering.}
We implement real-time rendering using deferred shading techniques~\cite{deering1988deferred}, based on the High Definition Render Pipeline (HDRP) in Unity. Specifically, we rasterize our 4D Gaussian sequence with extra PBR attributes, including base color, AO, normals, roughness, and depth maps, and store them in the GBuffer. This GBuffer is then integrated into the original forward transparent stage of HDRP for rendering semi-transparent objects. We also leverage shadow mapping from the HDRP rendering pipeline to perform shadow calculations for Gaussians under different light types. Our approach enables real-time rendering at 100 FPS in 1080P for volumetric videos.

\begin{figure*} [ht]
  \centering
  \includegraphics[width=\textwidth]{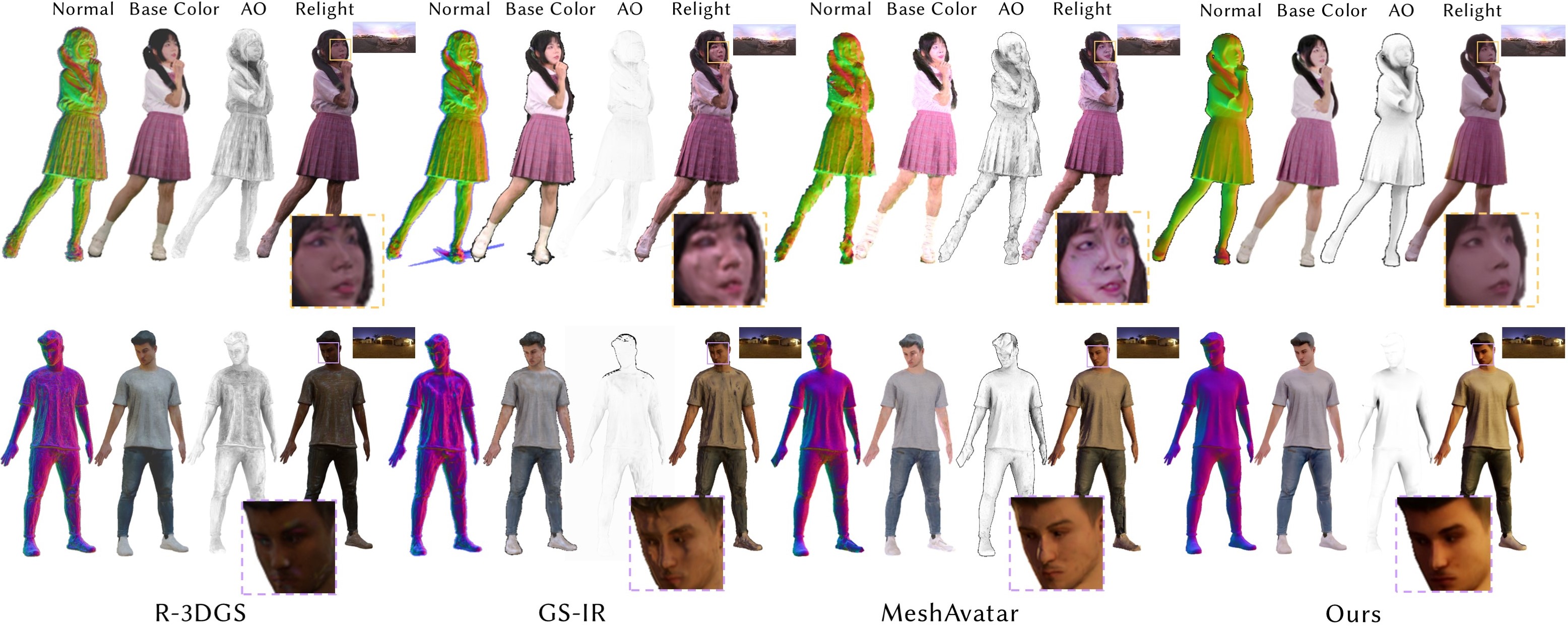}
  
  \captionof{figure}{Qualitative comparisons of our method against R-3DGS~\cite{gao2025relightable}, GS-IR~\cite{liang2024gs} and MeshAvatar~\cite{chen2024meshavatar}. Our method achieves the highest relighting quality.} 
  \vspace{-4pt}
  \label{fig_comparison1}
\end{figure*}

\paragraph{Offline Rendering.} 
For offline rendering, we sort the intersection points of the Gaussian during ray tracing and use alpha blending to determine the exact intersection with the entire Gaussian object, simultaneously acquiring the normal and material attributes at that point. This allows full compatibility with the widely used path tracing pipeline. To demonstrate the relighting quality of real-time rendering and offline rendering techniques, we showcase results for both under an environment map, as depicted in Fig.~\ref{fig_rendering}. Notably, the offline rendering effectively handles the shadows caused by occlusions in the Gaussian representations.
\section{Application}
\label{sec:application}

We perform relighting experiments using the recovered geometry, material, and illumination from our BEAM pipeline. As shown in Fig.~\ref{fig_moreres}, the results demonstrate that our method effectively handles diverse lighting conditions, producing photorealistic renderings with accurate material response.

In Fig.~\ref{fig_application}, we showcase real-time editing of relightable 4D Gaussian sequences within the Unity platform, including lighting adjustment and scene composition. Our pipeline integrates smoothly with conventional CG workflows, enabling efficient scene and lighting editing for artists and designers. With the support of a custom Unity plugin, our system provides instant rendering feedback, significantly enhancing the speed and flexibility of iterative design and optimization.

Leveraging Unity’s cross-platform support, we seamlessly deploy the edited 4D scenes to VR headsets for immersive rendering and interaction. Multiple human performances can be aligned and rendered together in immersive environments, allowing for compelling and realistic virtual experiences. Users can explore the scene freely, interact with lighting and assets in real time, and experience performances from novel viewpoints, making it a powerful tool for virtual production, digital exhibitions, and immersive storytelling.

\section{Experiments}
\label{sec:experiments}
To demonstrate our relighting capabilities, we capture 6 diverse human performances featuring detailed textures and challenging body motions, using an array of 81 Z-CAM cinema cameras at 3840×2160 resolution and 30 FPS, under an adequately illuminated dome environment.
Our pipeline is implemented based on 3DGS~\cite{gaussiansplatting} and trained on a single NVIDIA GeForce RTX 3090 GPU. Our method achieves a processing time of 12 minutes per frame for 4D Gaussians modeling and optimization, and 4 minutes for materials estimation and baking. The generated relightable 4D Gaussian sequences are fully compatible with VR platforms and CG engines, enhancing immersive experiences during playback and editing. 

\subsection{Comparison}
We compare our method with several state-of-the-art approaches. The static Gaussian relighting techniques, Relightable-3DGS~\cite{gao2025relightable} and GS-IR~\cite{liang2024gs}, reconstruct geometry frame by frame. The dynamic relighting method MeshAvatar~\cite{chen2024meshavatar} relies on an SMPL skeleton as a geometric proxy for lighting calculation. To ensure fairness, our comparison with MeshAvatar focuses on the relighting of reconstructed performance, rather than novel poses.
We also provide these methods with the same environment maps used by our method. 
As shown in Fig.~\ref{fig_comparison1}, the normals and AO decoupled by Relightable-3DGS are blurry, resulting in significant relighting artifacts. GS-IR struggles to reconstruct smooth normals and effectively separate the base color from other attributes. 
And MeshAvatar is constrained by mesh topology and SMPL prior, leading to reconstruction artifacts such as mesh distortion and surface tearing. These issues hinder the effectiveness of subsequent PBR-based disentanglement and relighting, resulting in suboptimal quality for high-frequency details.
In contrast, our method produces smooth normals and accurately decouples the AO and base color, enabling high-fidelity relighting results.
\begin{table}[t]
    \caption{
        Quantitative comparison with SOTA relighting methods on our synthetic dataset. \colorbox{best1}{Green} and \colorbox{best2}{yellow} cell colors indicate the best and the second-best results.
    }
    \label{tab:comparison_full}
    \centering

    \addtolength{\tabcolsep}{-2pt}
    \resizebox{\linewidth}{!}{

        \begin{tabular}{l|cccccccccc}  %
            \hline
            & \multicolumn{3}{c}{\textbf{AO}} & \multicolumn{3}{c}{\textbf{Base Color}} & \multicolumn{3}{c}{\textbf{Relighting}}                                                                                                                                 \\
            Method                              & PSNR$\uparrow$                          & SSIM$\uparrow$                          & LPIPS$\downarrow$ & PSNR$\uparrow$ & SSIM$\uparrow$ & LPIPS$\downarrow$ & PSNR$\uparrow$ & SSIM$\uparrow$ & LPIPS$\downarrow$ \\
            \hline
            
            R-3DGS         & \colorbox{best2}{20.67}      & 0.791   & \colorbox{best2}{0.307}         & 20.65         & 0.853       & 0.166       & \colorbox{best2}{24.66}     & 0.855      & \colorbox{best2}{0.105}            \\
            GS-IR      & 18.95     & \colorbox{best2}{0.850}    & 0.345    & 18.86     & 0.773      &  0.281    & 23.05      & \colorbox{best2}{0.858}  & 0.177     \\
            MeshAvatar    & 19.53     & 0.735       & 0.459    & \colorbox{best2}{21.04}     &  \colorbox{best2}{0.867}       & \colorbox{best2}{0.146}     & 23.93      & 0.857    & 0.178           \\
            Ours      & \colorbox{best1}{25.32}       & \colorbox{best1}{0.924}     & \colorbox{best1}{0.168}    & \colorbox{best1}{21.47}    & \colorbox{best1}{0.906}   & \colorbox{best1}{0.084}            & \colorbox{best1}{26.57}         & \colorbox{best1}{0.895}          & \colorbox{best1}{0.086}             \\
            \hline
        \end{tabular}
    }
    \addtolength{\tabcolsep}{2pt}
    \vspace{-20pt}
\end{table}

For quantitative comparison, evaluations are conducted on synthetic data to generate ground truth images under predefined lighting conditions. We use Blender to simulate the similar capture perspectives of our dome system and render human meshes from RenderPeople\cite{RenderPeople} into corresponding viewpoints using the CYCLES engine.
Reconstruction quality is assessed using three widely used quantitative metrics: PSNR, SSIM, and LPIPS. To ensure a fair and precise comparison, we compute the metrics for AO, base color, and relighting results across two synthetic sequences, each consisting of 150 frames. In addition, we compute these metrics within the bounding box of the human region. As shown in Tab.~\ref{tab:comparison_full}, our method surpasses the other techniques in all metrics evaluated.

\subsection{Evaluation}
\begin{figure}[h]
  \includegraphics[width=\columnwidth]{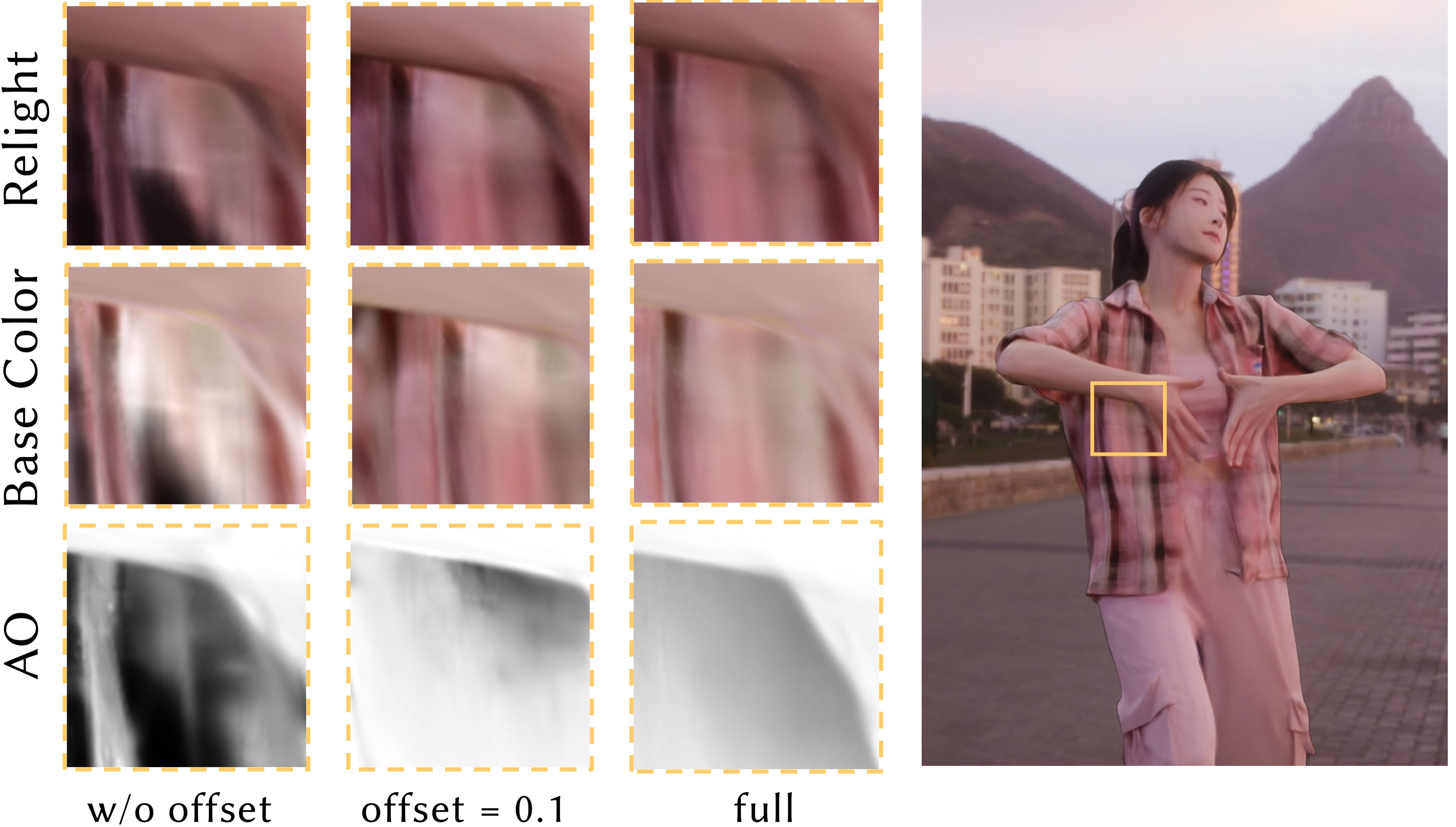}
  \captionof{figure}{Ablation on offset variable for materials decomposition. Our full model provides accurate AO and base color representations. }
  \label{fig_ablation_study}
  \vspace{-4pt}
\end{figure}

\paragraph{Materials Decomposition.}
We conduct a qualitative ablation study on materials decomposition to examine the impact of different variables on AO, base color, and the final relighting results. For the offset variable in the Gaussian surface estimation during ray tracing, we present the results without the offset and with an excessively large offset (0.1), in the first and second columns of Fig.~\ref{fig_ablation_study}. We observe that without an offset, AO and base color contained many black artifacts, and the Gaussian surface estimation suffered significant degradation. On the other hand, an offset of 0.1 led to overestimation, causing incorrect lighting decoupling and producing overly bright artifacts. Instead, an offset of 0.02 is applied for human data in the general case.

\begin{figure}[h]
  \includegraphics[width=\columnwidth]{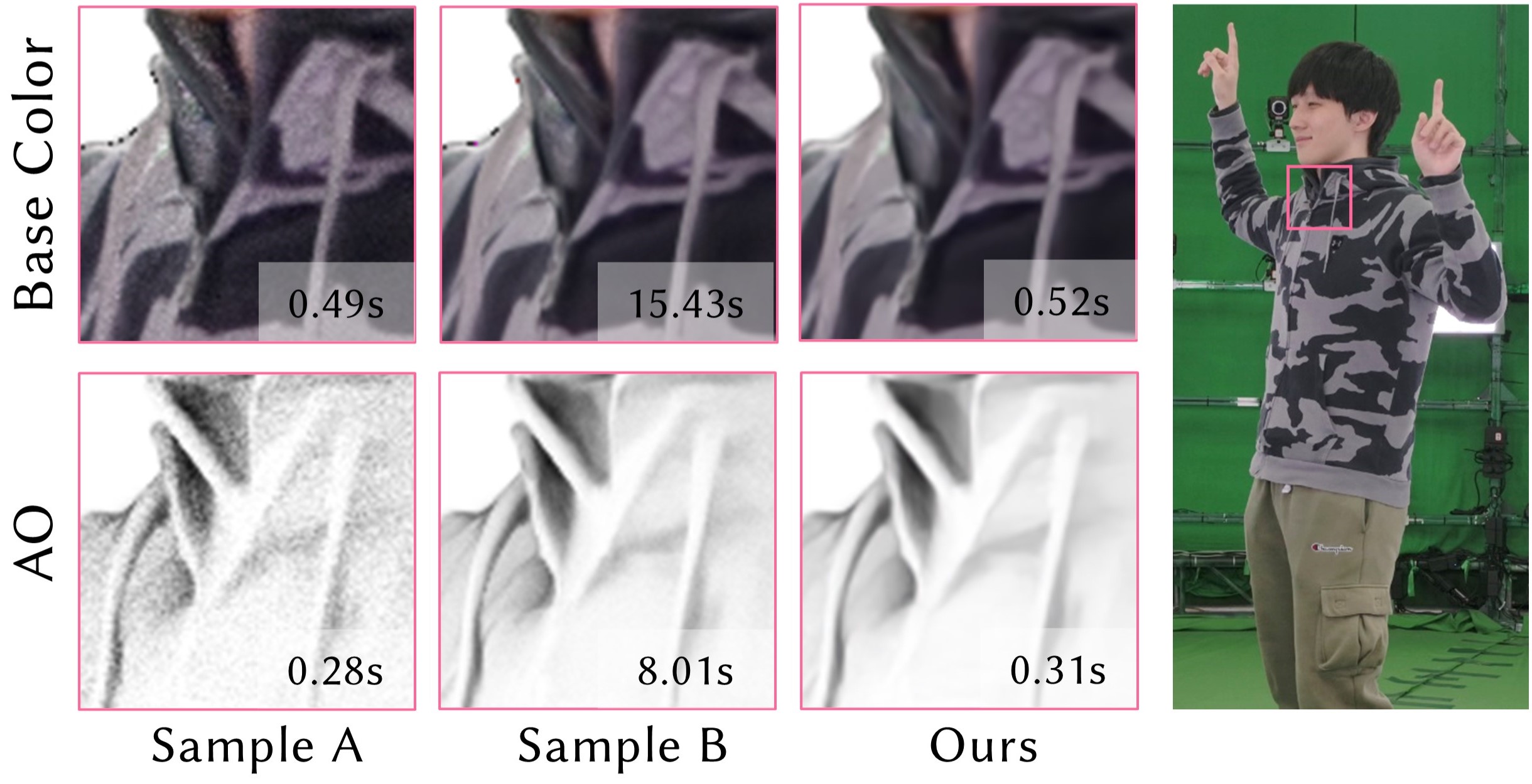}
  \captionof{figure}{Ablation on sampling strategies for material decomposition. Rendering time for each image is annotated in the zoomed-in views. }
  \label{fig_ablation_study_sample}
  \vspace{-8pt}
\end{figure}

In addition, we evaluate the sampling strategy used in the Gaussian sampler for disentangling base color and AO. Specifically, we compare the effects of sampling count and the denoising component. We present results under three configurations: (1) Sample A: low sample counts (50 spp for AO, 100 spp for base color) without denoising, (2) Sample B:  high sample counts (1000 spp for AO, 2000 spp for base color) without denoising, and (3) our adopted strategy using low sample counts with denoising. As shown in Fig.~\ref{fig_ablation_study_sample}, our sampling strategy achieves a favorable balance between rendering quality and computation time, whereas alternative strategies either suffer from noisy outputs or incur significantly higher computational costs.

\begin{figure}[t]
  \includegraphics[width=\columnwidth]{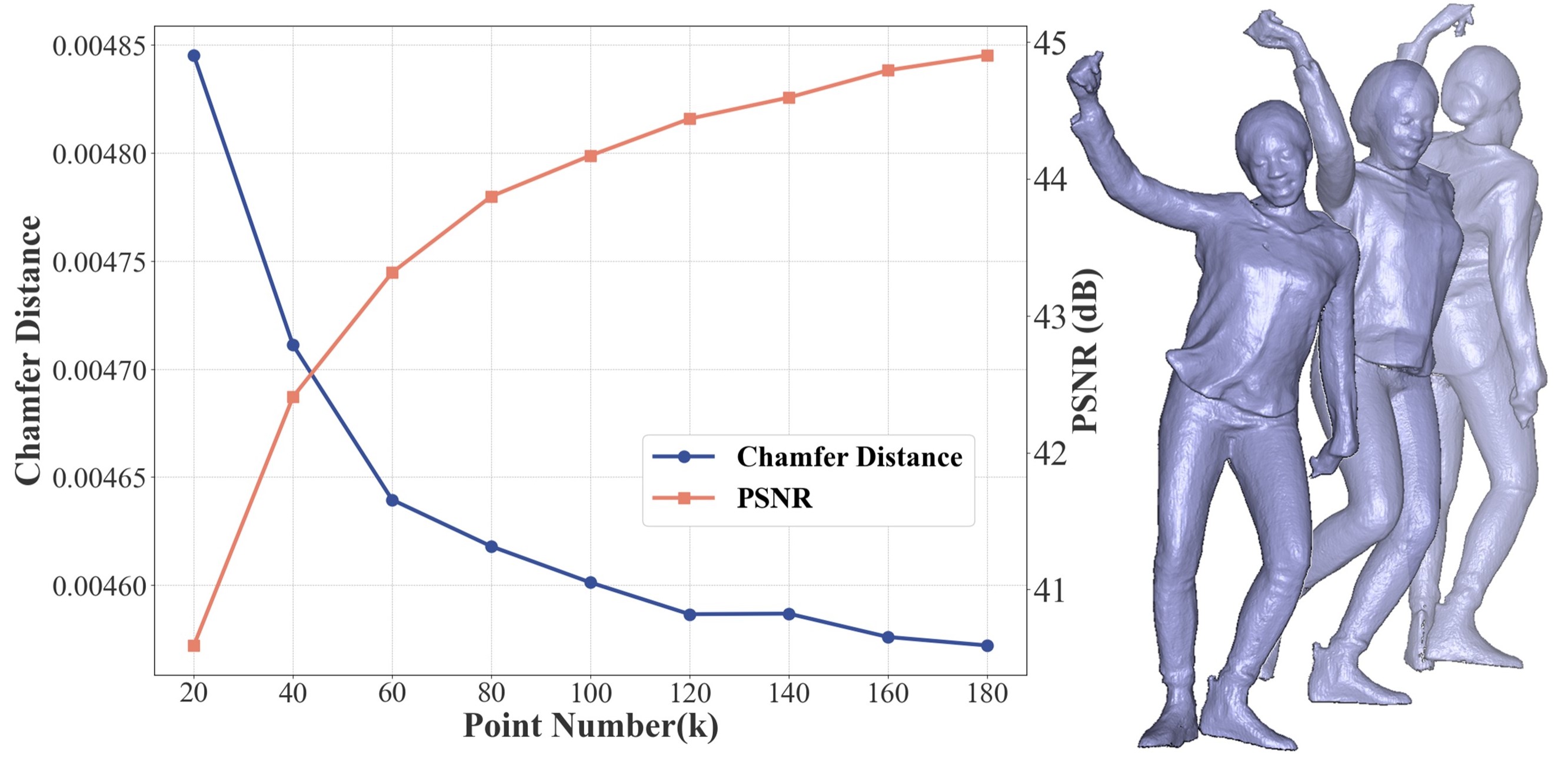}
  \captionof{figure}{Ablation study on the number of geometry-aware Gaussians. With $\sim140,000$ Gaussians, BEAM achieves high rendering quality and geometric reconstruction accuracy, while ensuring efficient training and rendering.} 
  \label{fig_ablation_study_point}
  \vspace{-12pt}
\end{figure}

\paragraph{The Number of Gaussians.}
We evaluate the impact of varying the number of 4D Gaussians on both rendering quality and mesh quality across 100 frames in our virtual dataset, using PSNR and Chamfer Distance, respectively. As shown in Fig.~\ref{fig_ablation_study_point}, we visualize how key metrics vary with the number of Gaussians, along with the mesh extracted from the Gaussian sequence at 140,000 points. We observe that using approximately 140,000 geometry-aware Gaussians achieves an optimal balance between rendering quality and geometric fidelity, while avoiding unnecessary redundancy. This point count ensures efficient training and is well-suited for applications such as VR and AR.

\paragraph{The Number of Cameras.}
To assess our method's robustness to sparse input views, we conduct an ablation study on the number of input camera views. While our proposed method was trained with 50 camera views on synthetic data, we also evaluate it on uniformly selected subsets of 20, 30, and 40 views. As shown in Tab.~\ref{tab:ablation_cam}, our method demonstrates stable performance even with fewer cameras, showing its potential for generalization to sparse-view tasks.

\begin{table}[t]
    \centering
    \caption{
        Quantitative ablation on the number of input camera views.
    }
    \resizebox{\linewidth}{!}{
        \begin{tabular}{l|cccccc}
        \hline
        \textbf{Methods} & \multicolumn{3}{c}{\textbf{Render}} & \multicolumn{3}{c}{\textbf{Relighting}} \\
        \cline{2-7}
        & \textbf{PSNR $\uparrow$} & \textbf{SSIM $\uparrow$} & \textbf{LPIPS $\downarrow$} & \textbf{PSNR $\uparrow$} & \textbf{SSIM $\uparrow$} & \textbf{LPIPS $\downarrow$} \\
        \hline
        20 cams & 34.56 & 0.935 & 0.0758 & 24.47 & 0.850 & 0.1056 \\
        30 cams & 38.40 & 0.964 & 0.0598 & 25.07 & 0.854 & 0.0957 \\
        40 cams & 39.73 & 0.975 & 0.0494 & 25.36 & 0.858 & 0.0924 \\
        \hline
        Ours & 42.73 & 0.980 & 0.0444 & 26.21 & 0.861 & 0.0892 \\
        \hline
        \end{tabular}
    }
    \label{tab:ablation_cam}
    \vspace{-8pt}
    
\end{table}

\subsection{User Study}
We conduct a user study to evaluate the temporal reconstruction quality and normal consistency of our 4D Gaussians. For each method, we prepare visualizations consisting of normal maps and relighting results rendered under a consistent skybox environment to highlight illumination response. Specifically, for GS-IR and R-3DGS, Gaussian models are trained per frame over 200 frames, and both normal maps and relighting results are rendered under a shared HDR skybox. For our method and MeshAvatar, normal maps and relighting are generated from a dynamic 200-frame Gaussian sequence using the same skybox. We present the results to 30 users and ask them to select the most visually realistic rendering under novel lighting.
In terms of temporal reconstruction quality and relighting quality, 95.65\% of users prefer our method, while 87\% choose our approach for normal consistency.
These preference results clearly indicate a significant advantage of our method over the competing approaches, demonstrating its superior performance in both aspects.

\section{Conclusion}
\label{sec:conclusion}
\paragraph{Limitations.}
Although our method achieves high-quality immersive rendering, there are some limitations. 
First, we approximate the rendering equation to obtain the 2D material maps, which introduces errors in our decoupled material results and does not accurately reflect the real physical world.
Future work may address this issue by incorporating large models like video generation. Furthermore, our approach, focusing on relighting for a dynamic reconstruction sequence, does not support pose-driven animation or the generation of new poses. Future work will focus on optimizing these aspects to improve robustness and applicability.

We have presented a Gaussian-based approach for reconstructing detailed geometry and PBR materials to produce relightable volumetric videos. We employ a coarse-to-fine training strategy and effective geometric constraints to accurately model the dynamic geometry of 4D Gaussians. Additionally, we decouple PBR materials by using ray tracing to compute the lighting effects and obtain base color and AO maps, while leveraging generative methods to infer roughness. These materials are then baked into the corresponding attributes of the Gaussians. With deferred shading and ray tracing techniques, our Gaussian sequence supports both efficient real-time rendering and more realistic offline rendering. Experimental results demonstrate the advantages of our approach in generating high-quality dynamic normal maps and material decomposition, as well as its relightability under a variety of lighting conditions. Our method is highly compatible with traditional CG engines, offering significant potential for enhancing rendering realism and flexibility, enabling users to immerse themselves in and interact with dynamic, relightable volumetric worlds.
\begin{acks}
This work was supported by the National Key R$\&$D Program of China (2022YFF0902301), Shanghai Local College Capacity Building Program (22010502800), and Shanghai Frontiers Science Center of Human-centered AI (ShangHAI).
\end{acks}


\bibliographystyle{ACM-Reference-Format}
\bibliography{sample-base}

\end{document}